\documentclass[11pt]{article}
\usepackage{epsf}
\begin{document}

\noindent
{\LARGE {\bf Decomposing the stock market intraday \\ dynamics}}

\vspace{0.1cm}

\begin{center}

J. Kwapie\'n$^1$, S. Dro\.zd\.z$^{1,2}$, F. Gr\"ummer$^3$, F. Ruf$^4$,
J. Speth$^3$

\vspace{0.2cm}

{\it
$^1$Institute of Nuclear Physics, Radzikowskiego 152,
31--342 Krak\'ow, Poland

$^2$Institute of Physics, University of Rzesz\'ow,
35--310 Rzesz\'ow, Poland

$^3$Institute f\"ur Kernphysik, FZ J\"ulich, D--52425 J\"ulich, Germany

$^4$West LB International S.A., 32--34 bd Grande--Duchesse Charlotte,
L--2014
Luxembourg}

\end{center}

\centerline{\bf Abstract}

{\small{\it
The correlation matrix formalism is used to study temporal aspects of
the stock market evolution. This formalism allows to decompose the financial
dynamics into noise as well as into some coherent repeatable intraday
structures. The present study is based on the high-frequency Deutsche
Aktienindex (DAX) data over the time period between November 1997 and
September 1999, and makes use of both, the corresponding returns as well
as volatility variations. One principal conclusion is that a bulk of the
stock market dynamics is governed by the uncorrelated noise-like
processes. There exists however a small number of components of coherent
short term repeatable structures in fluctuations that may generate some
memory effects seen in the standard autocorrelation function analysis.
Laws that govern fluctuations associated with those various components are
different, which indicates an extremely complex character of the financial 
fluctuations.}}

\section{Introduction}

One of the great challenges of econophysics is to properly quantify and,
following this, to explain the nature of financial correlations and
fluctuations. The efficient market hypothesis~\cite{Samu} implies that
they are dominated by noise. Indeed, the spectrum of the 
correlation matrix accounting for correlations among the stock market
companies agrees very well~\cite{Lalo,Pler,Dro1} with the universal
predictions of random matrix theory~\cite{Wign,Meht}.
Locations of extreme eigenvalues differ however from these predictions
and thus identify certain system-specific, non-random properties such
as collectivity. In addition, these former properties turn 
out~\cite{Dro1,Dro2} to depend on time reflecting a competitive character 
of the financial dynamics.

\begin{figure}
\hspace{1.5cm}
\epsfxsize 9.0cm
\epsfysize 6.0cm
\epsffile{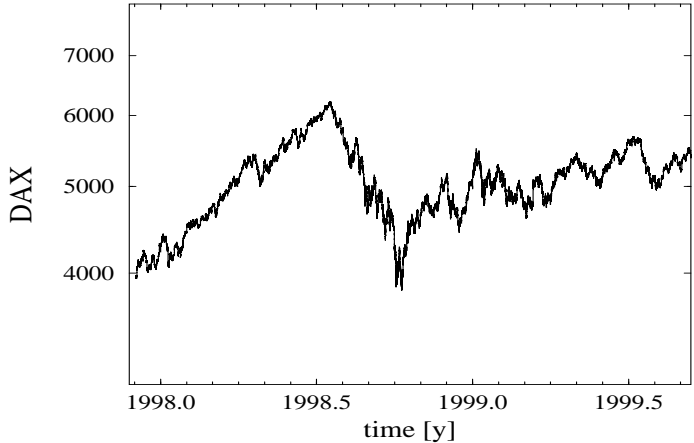}
\caption{The Deutsche Aktienindex (DAX) in the calendar period November
28, 1997 -- September 17, 1999.}
\label{fig:fig1}
\end{figure}

The character of the financial time correlations is however very complex, 
still poorly understood and many related issues remain puzzling.     
The autocorrelation function of the financial time series, for instance, 
drops down to zero within few minutes which is interpreted as a time horizon 
of the market inefficiency~\cite{Camp}. 
At the same time, however, the correlations in volatility are significantly 
positive over the time intervals longer by many orders of magnitude.
The fat-tailed return distributions seem to be not L\'evy
stable~\cite{Stan} on short time scales, but on longer time scales it
appears difficult to identify their convergence to 
a Gaussian as expected from the central limit theorem. In addressing this
sort of issues below we use the concept of the correlation matrix whose
entries are constructed from the time series of price changes
representing the consecutive trading days. The method focuses then
entirely on the time correlations and their potential existence can 
parallelly be detected on various time scales.
Analogous methodology has already been successfully applied~\cite{Kwap} 
to extract from noise some repeatable structures in the brain sensory
response, and its somewhat similar variant, the correlation matrix of the
delay matrix, to study the business cycles of economics~\cite{Orme}.
The present study is an extension of our recent work~\cite{Dro3} and is based  
on an example of high-frequency (15 s) recordings~\cite{data}.
As it can be seen from Fig.~\ref{fig:fig1}, 
this is an interesting period which comprises
the whole richness of the stock market dynamics like strong increases 
and decreases, and even a clearly identifiable hierarchy of the
log-periodic structures~\cite{Dro4}.

\section{Definiton of correlations}

In the present application the entries of the correlation matrix 
are constructed from the time-series $g_{\alpha}(t_i)$ of normalized 
price returns representing the consecutive trading days labelled by
$\alpha$. Starting from the original price time-series $x_{\alpha}(t)$ 
these are defined as 
\begin{equation}
g_{\alpha}(t_i) = {G_{\alpha}(t_i) - \langle G_{\alpha}(t_i) \rangle_t
\over \sigma(G_{\alpha})} \ , \quad
\sigma(G_{\alpha}) = \sqrt{\langle G_{\alpha}^2(t) \rangle_t -
\langle G_{\alpha}(t) \rangle_t^2} \ ,
\label{eq:normret}
\end{equation}
with
\begin{equation}
G_{\alpha}(t_i) = \ln x_{\alpha}(t_i+\tau) - \ln x_{\alpha}(t_i)
\simeq {x_{\alpha}(t_i+\tau) - x_{\alpha}(t_i) \over x_{\alpha}(t_i)}
\ ,
\label{eq:ret}
\end{equation}
where $\tau$ is the time-lag and $\langle\ldots\rangle_t$ denotes
averaging over time. 

The result is $N$ time series $g_{\alpha}(t_i)$ of length $T$ (the number
of records during the day) {\it i.e.} an $N \times T$ matrix $\bf M$. The
correlation matrix can then be defined as 
\begin{equation}
{\bf C} = (1 / T) \ {\bf M} {\bf M}^{\bf T}. 
\label{eq:C}
\end{equation}
Its entries $C_{\alpha,\alpha'}$ are thus labelled by the pairs of
different days. By diagonalizing $\bf C$ 
\begin{equation}
{\bf C} {\bf v}^k = \lambda_k {\bf v}^k,
\label{eq:eig}
\end{equation}
one obtains the eigenvalues $\lambda_k$ $(k=1,...,N)$ and the corresponding
eigenvectors ${\bf v}^k = \{ v^k_{\alpha} \}$.

A useful null hypothesis is provided by the limiting case of entirely 
random correlations. In this case the density of
eigenvalues $\rho_C(\lambda)$ defined as 
\begin{equation}
\rho_C(\lambda) = {1 \over N} {{dn(\lambda)} \over {d \lambda}},
\label{eq:rho}
\end{equation}
where $n(\lambda)$ is the number of eigenvalues of $\bf C$ less than $\lambda$,
is known analytically~\cite{Edel}, and reads
\begin{eqnarray}
\hspace{2.5cm}
\label{eq:rho1}
\rho_C(\lambda) = {Q \over {2 \pi \sigma^2}} 
{\sqrt{ (\lambda_{max} - \lambda) (\lambda - \lambda_{min})} \over
{\lambda}}, \\
\centering
\lambda^{max}_{min} = \sigma^2 (1 + 1/Q \pm 2 \sqrt{1/Q}),
\nonumber
\end{eqnarray}
with $\lambda_{min} \le \lambda \le \lambda_{max}$, $Q=T/N \ge 1$, and
where $\sigma^2$ is equal to the variance of the time series which
in our case equals unity.

\section {DAX time-correlations}

As mentioned above our related study is based on the DAX 
recordings with the frequency of 15 s during the period between 
November 28th, 1997 and September 17th, 1999. 
After this last date the DAX was traded significantly longer during 
the trading day. By taking the DAX intraday
15 s variation between the trading time 9:03 and 17:10 which corresponds to
$T=1948$, and rejecting several days with incomplete recordings, 
one then obtains $N=451$ complete and equivalent time series
representing different trading days during this calendar period. 
Using this set of data we then construct the $451 \times 451$ matrix $\bf C$.

\begin{figure}
\epsfxsize 8.0cm
\hspace{2.1cm}
\epsffile{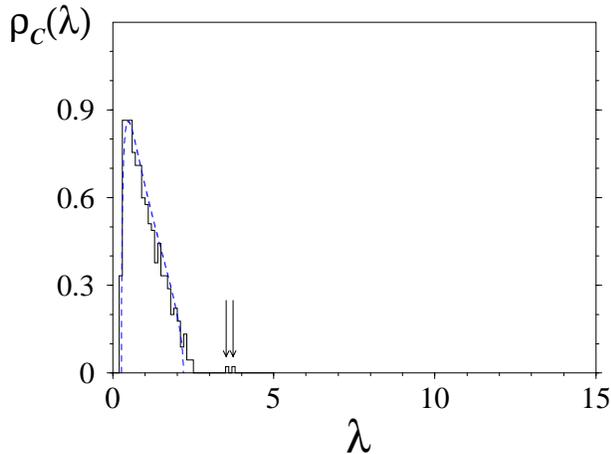}
\caption{The probability density (histogram) of the eigenvalues of the
correlation matrix $\bf C$ calculated from the DAX time series of 15 s
returns during the calendar period November 28, 1997 -- September 17,
1999. The null hypothesis of purely random correlations formulated in
terms of Eq.~(\ref{eq:rho1}) is indicated by the dashed line.}
\label{fig:fig2}
\end{figure}

One characteristic of interest is the structure of eigenspectrum.
The resulting probability density of eigenvalues, shown in
Fig.~\ref{fig:fig2},
displays a very interesting structure. 
There exist two almost degenerate eigenvalues visibly repelled from 
the bulk of the spectrum, i.e., well above  
$\lambda_{max}$ (for $Q= 1948/451$, $\lambda_{max} \approx 2.19$) 
which indicates that the dynamics develops
certain time specific repeatable structures in the intraday trading.  
The bulk of the spectrum, however, agrees remarkably well with the bounds 
prescribed by purely random correlations.
This indicates that the statistical neighbouring recordings in our time 
series of 15 s DAX returns share essentially no common information. 

\begin{figure}
\epsfxsize 8.0cm
\hspace{2.1cm}
\epsffile{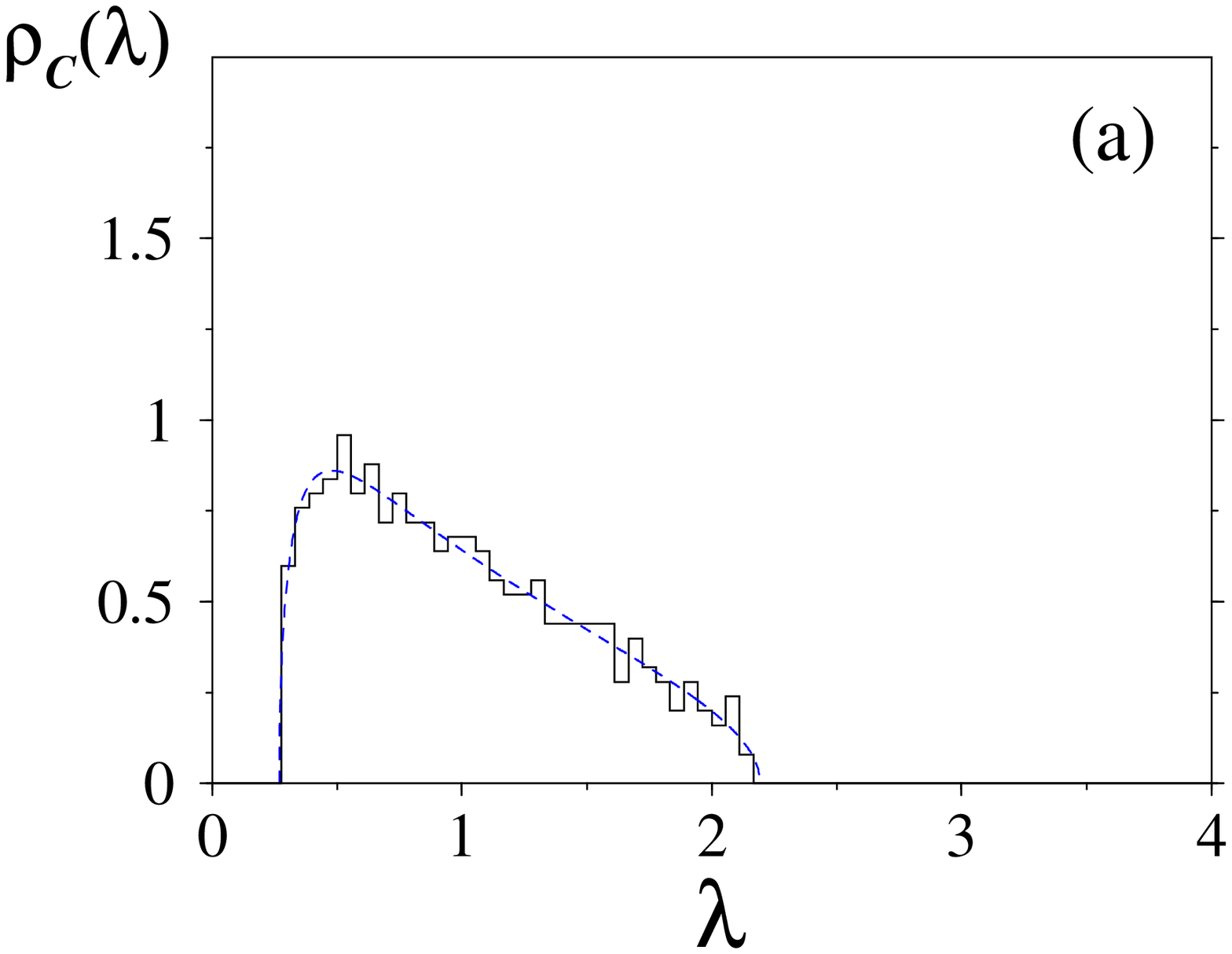}

\vspace{0.0cm}
\epsfxsize 8.0cm
\hspace{2.1cm}
\epsffile{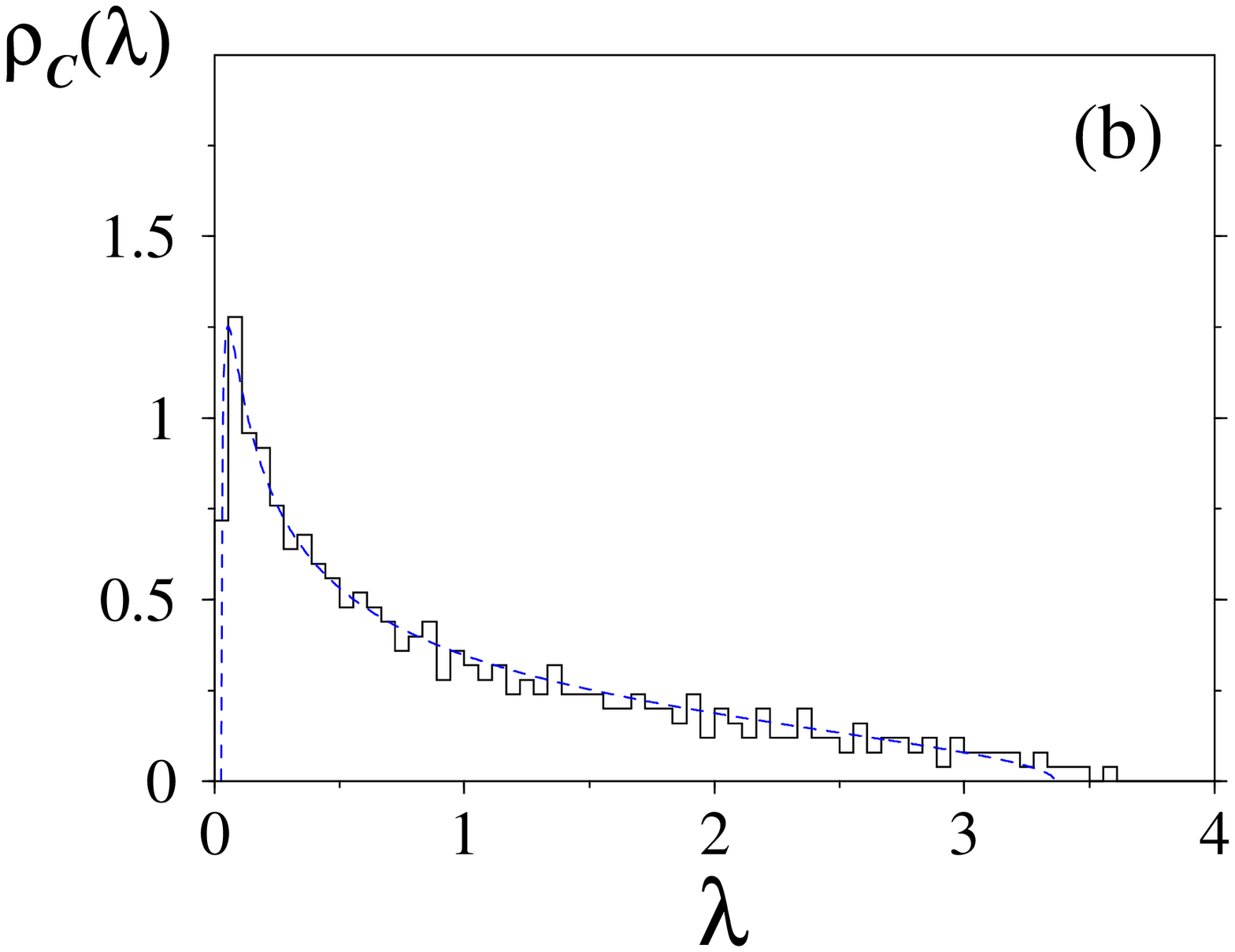}
\caption{(a) The probability density (histogram) of eigenvalues of the
correlation matrix $\bf C$ calculated from $N=451$ series of length
$T=1948$ of the Gaussian distributed uncorrelated random numbers versus
the corresponding null hypothesis (dashed line) formulated in terms of
Eq.~(\ref{eq:rho1}). (b) The same as (a) but here only every third
original random number in each series is retained. The removed pairs of
numbers are replaced by the new ones which are functionally dependent on
the neighbouring two original numbers. The dashed line indicates the
result of Eq.~(\ref{eq:rho1}) for $N$ series of length $T/3$.}
\label{fig:fig3}
\end{figure}

\begin{figure}
\epsfxsize 8.0cm
\hspace{2.1cm}
\epsffile{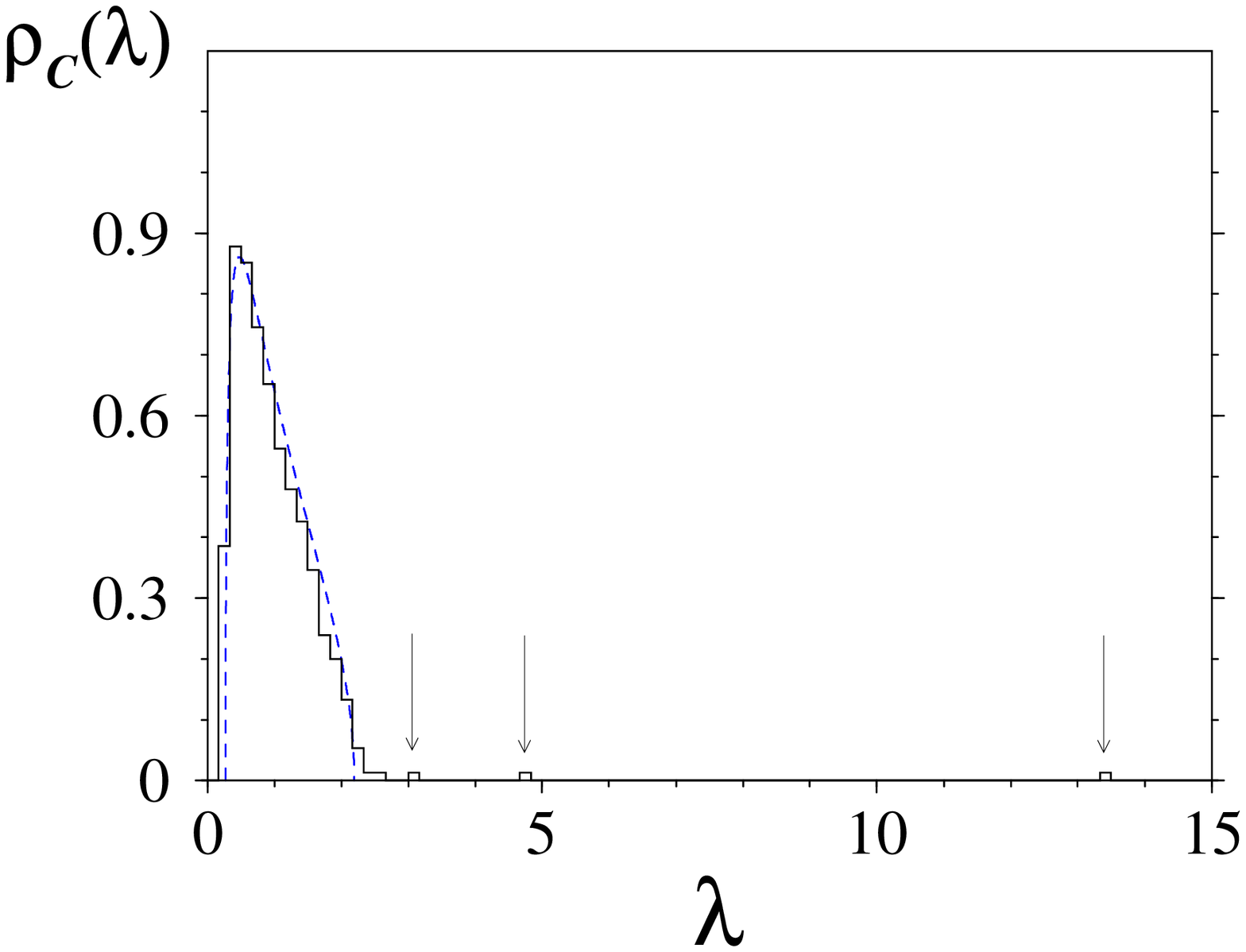}
\caption{The same as in Fig.~\ref{fig:fig2} but now the volatilities are
used instead of returns.}
\label{fig:fig4}
\end{figure}

A significance of this result can be evaluated by the following numerical
experiment. From a Gaussian distribution we draw $N=451$ series 
$x_n(i)$ $(n=1,...,N)$ of random numbers of length $T=1948$ $(i=1,...,T)$  
and determine the spectrum of the resulting correlation matrix.
The result (histogram) versus the corresponding theoretical result
expressed by the Eq.~(\ref{eq:rho1}) is shown in the upper part of
Fig.~\ref{fig:fig3}. As expected, the agreement is unqestionable. In the
second step, in each previous series we retain only every third number,
e.g., $x_n(1), x_n(4), ...$ This omission
is compensated by insertion between every two remaining original numbers,
say $x_n(i)$ and $x_n(i+3)$, the two new $x_n(i+1), x_n(i+2)$ numbers 
such that they are functionally (here linearly) dependent on $x_n(i)$ and
$x_n(i+3)$. The net result is the same number $N$ of series of the same
length $T$ as before, thus $Q=T/N$ formally remains unchanged.
The structure of eigenspectrum of the corresponding correlation matrix, 
which is shown in the lower part of 
Fig.~\ref{fig:fig3},
changes however completely. In fact, it now perfectly agrees with the
theoretical formula of Eq.~(\ref{eq:rho1}) but for the three times shorter
$(T/3)$ series, i.e., it nicely reflects a real information content.
From this we can conclude that a
common information shared by neighbouring events in our DAX time series
is basically null and that a whole nonrandomness can be 
associated with the two largest eigenvalues.

It is also interesting to perform an analogous study of the volatility 
correlations. This corresponds to replacing $G_{\alpha}(t_i)$ in
Eq.~(\ref{eq:normret}) by $\vert G_{\alpha}(t_i) \vert$, for instance.
The structure of eigenspectrum of the resulting correlation matrix is
shown in Fig.~\ref{fig:fig4}. Surprisingly, even in this case the bulk
of the spectrum is consistent with purely random correlations.
As compared to Fig.~\ref{fig:fig2}, one can now identify however three 
outlying eigenvalues and the largest of them is repelled
significantly higher, as far up as 13.3. 

\begin{figure}
\epsfxsize 8.0cm
\hspace{2.0cm}
\epsffile{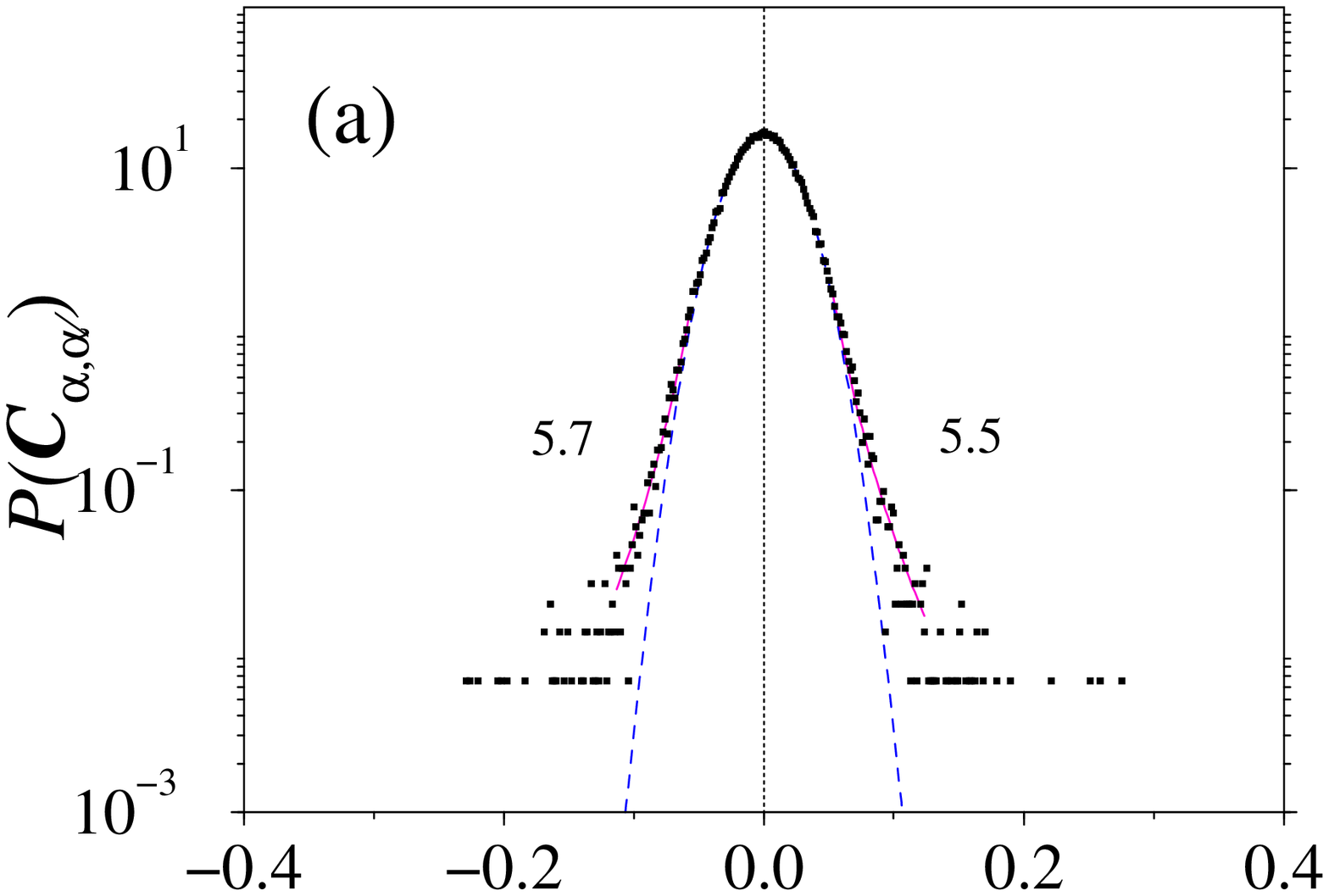}

\vspace{0.0cm}
\epsfxsize 8.0cm
\hspace{2.0cm}
\epsffile{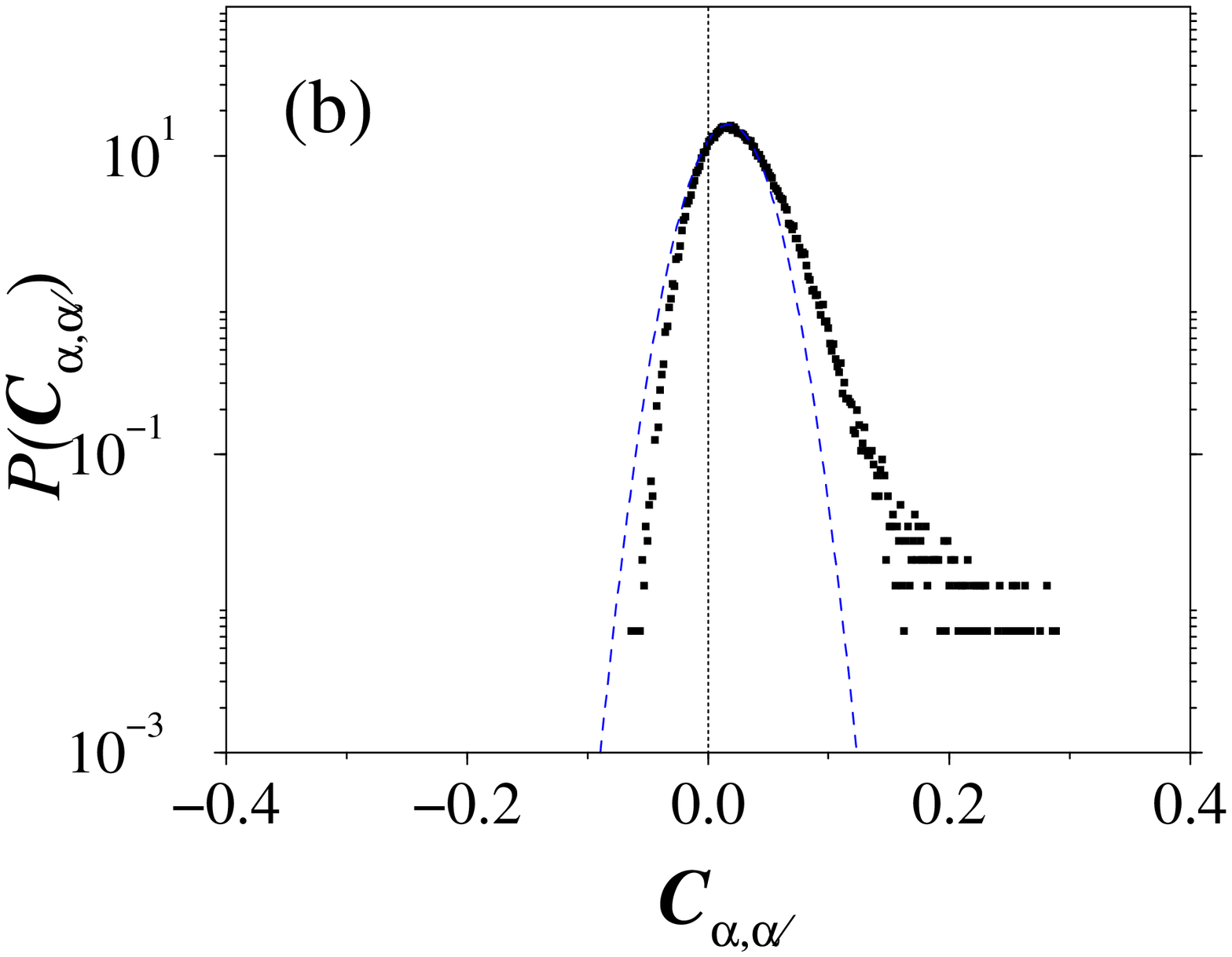}
\caption{Distribution of matrix elements $C_{\alpha,\alpha'}$ of the $N
\times N$ ($N=451$) correlation matrix $\bf C$ calculated from the 15 s
frequency DAX variation during the intraday trading time 9:03--17:10.
$\alpha$ labels the different trading days during the calendar period
December 28, 1997 -- September 17, 1999. The upper (a) part corresponds to
the time series of returns and the lower (b) part to the time series of
volatilities. The solid lines in (a) indicate the power law fits to the
tails of the distribution, while the dashed one represents a Gaussian
best fit. The numbers in (a) reflect the corresponding scaling indices.}
\label{fig:fig5}
\end{figure}

The structure of eigenspectrum of a matrix is expected to be
related~\cite{Dro5,Dro6} to the distribution of its elements. 
For this reason in Fig.~\ref{fig:fig5} we show the
distributions of such elements of $\bf C$ corresponding to the above
specified procedure for our two cases under consideration. 
The upper part of this figure corresponds to the returns time series
and the lower part to the volatility time series.
In the first case this distribution is symmetric with respect
to zero, a Gaussian like (dashed line) on the level of small matrix
elements, but sizably thicker than a Gaussian on the level of large matrix
elements, where a power law with the index of about 5.5 - 5.7 
(which however is far beyond the L\'evy stable
regime as consistent with the distribution of returns) 
provides a reasonable representation. It is these tails which generate
the two largest eigenvalues seen in Fig.~\ref{fig:fig2}.
The volatility correlation matrix, on the other hand, reveals a somewhat
different distribution. First of all, the center of this distribution
is shifted towards positive values and this is responsible for the largest 
eigenvalue. Secondly, this distribution is asymmetric. 
This originates from the fact that the volatility fluctuations are stongly 
asymmetric relative to their average value. The slope on the right hand
side cannot be here reliably measured in terms of a single power law, but
its even smaller value as compared to the previous case (of returns) is
evident. On the other hand, on the negative side the distribution drops
down faster than a Gaussian and, therefore, the separation between the two
remaining large eigenvalues is significantly more pronounced than of their
returns counterparts (two largest ones) from Fig.~\ref{fig:fig2}.

\begin{figure}
\epsfxsize 8.0cm
\hspace{1.7cm}
\epsffile{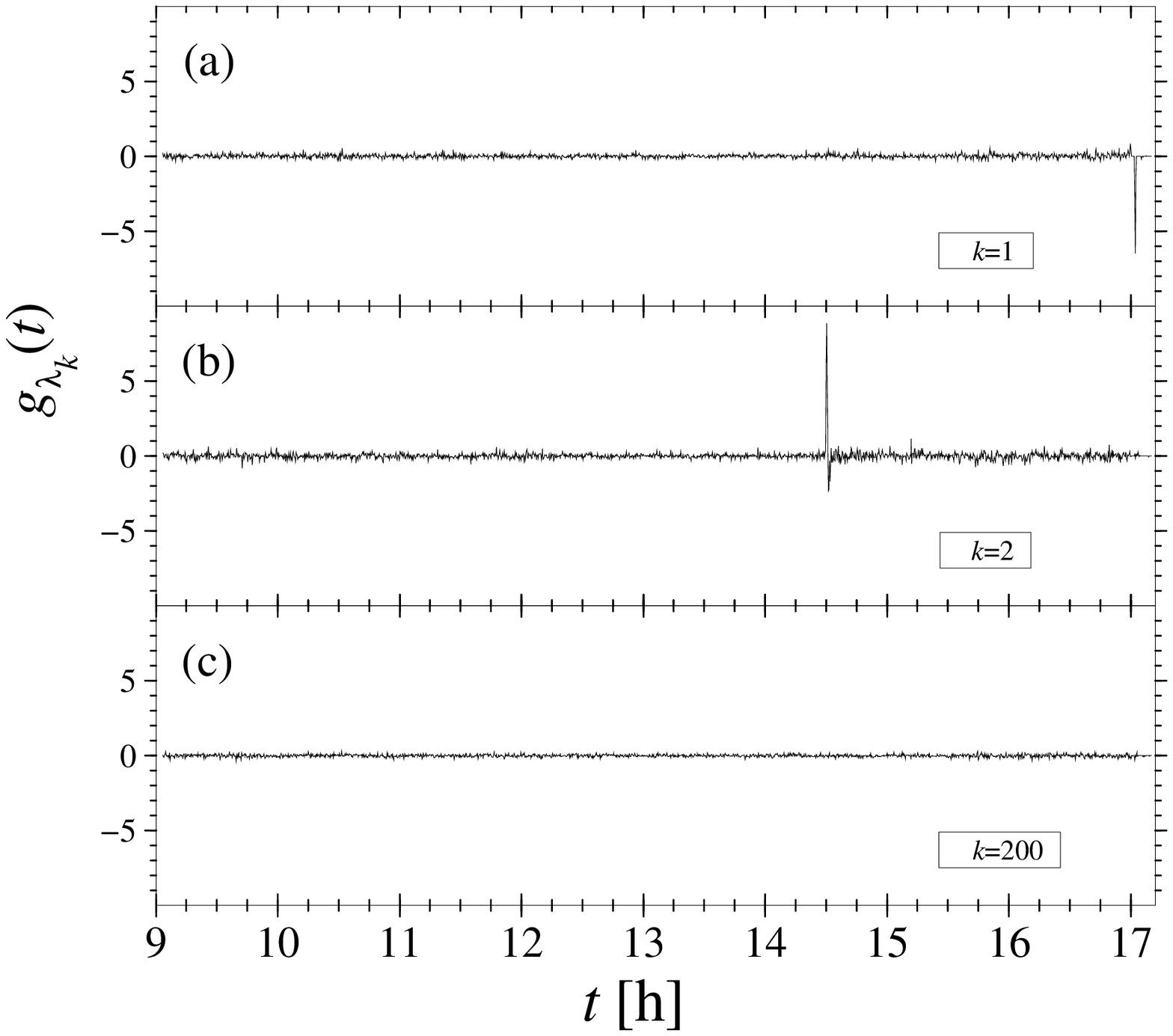}
\caption{The superposed time series of normalized returns calculated
according to eq.~(\ref{eq:gs}) for $k=1$ (a), $k=2$ (b) and $k=200$ (c).}
\label{fig:fig6}
\end{figure}

\begin{figure}
\epsfxsize 8.0cm
\hspace{1.7cm}
\epsffile{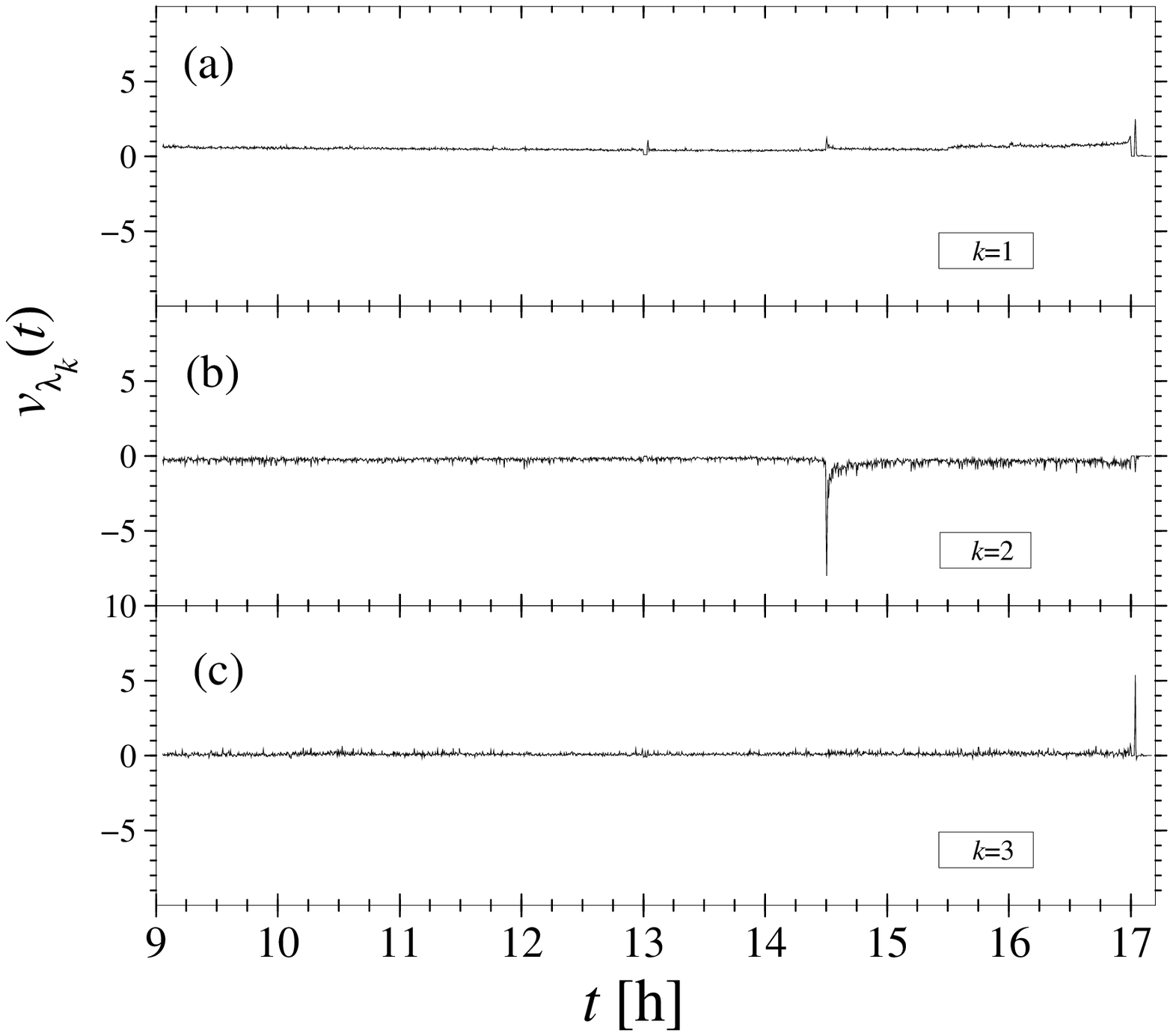}
\caption{The superposed time series of normalized volatilities calculated
according to eq.~(\ref{eq:nus}) for $k=1$ (a), $k=2$ (b) and $k=3$ (c).}
\label{fig:fig7}
\end{figure}

In quantifying the differences among the eigenvectors it is instructive 
to look at the superposed time series of normalized returns.
One possibility adopted here reads: 
\begin{equation}
g_{{\lambda}_k}(t_i) = \sum_{\alpha=1}^N {\rm sign}(v^k_{\alpha}) 
\vert v^k_{\alpha} \vert^2 g_{\alpha}(t_i).
\label{eq:gs}
\end{equation}
In this definition $\vert v^k_{\alpha} \vert^2$ is used instead of
$v^k_{\alpha}$ for the reason of preserving normalization and the sign of
$v^k_{\alpha}$ in order not to destroy any possible coherence among the
original signals. A collection of such superposed time series of returns 
for $k=1$, 2 and 200 is shown in Fig.~\ref{fig:fig6}. 
As it is illustrated by an example of $k=200$, for a statistical value of
$k$ such a superposed signal develops  
no coherent structures and $g_{{\lambda}_k}(t_i)$ basically does not 
differ from a simple average. 
The first two differ however
significantly and indicate the existence of the very pronounced,
up to almost ten times of the mean standard deviations of the original
time series, repeatable structures at the
well defined instants of time through many days. As it is clearly seen, 
the two collective signals correspond to two disconnected and well determined 
periods of an enhanced synchronous market activity. 
The first $(k=1)$ of them corresponds to the period just before 
closing in Frankfurt during the time interval considered here,
and the second one $(k=2)$ to the period immediately after 14:30,
which reflects the DAX response to the 
North-American financial news release exactly at this time.
It is also very interesting to see that in the first case (before closing)
the coherent burst of activity expressed by $g_{\lambda_1}(t)$ 
is oriented to the negative values while in the second case (just after 14:30) 
$(g_{\lambda_2}(t))$ it points predominantly to the positive values.  
Surprisingly, the DAX response to the Wall Street opening at 15:30
develops no visible synchronous structure in neither of the eigenstates.   

\begin{figure}
\epsfxsize 8.0cm
\hspace{2.1cm}
\epsffile{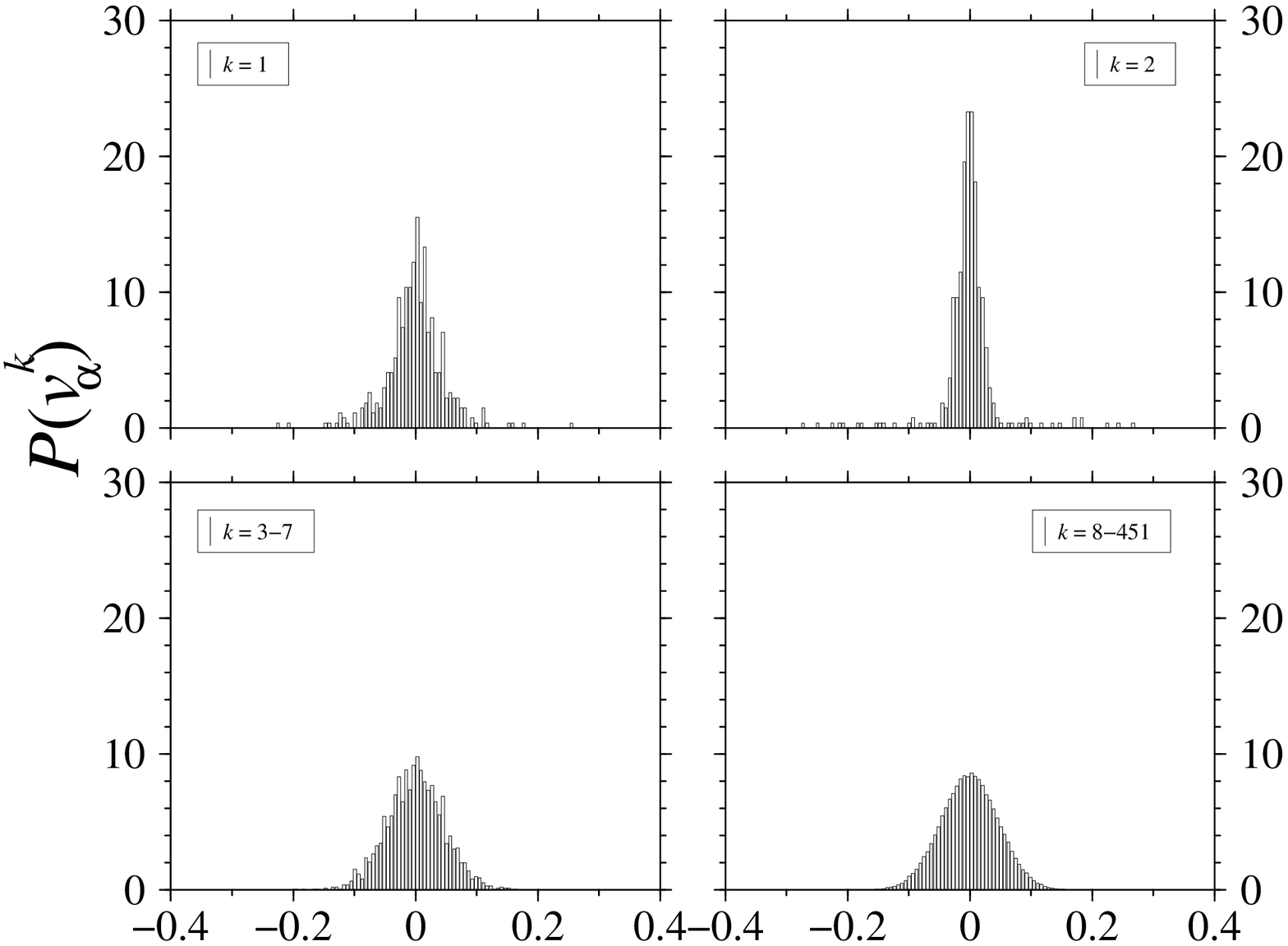}

\vspace{0.0cm}
\epsfxsize 8.0cm
\hspace{2.1cm}
\epsffile{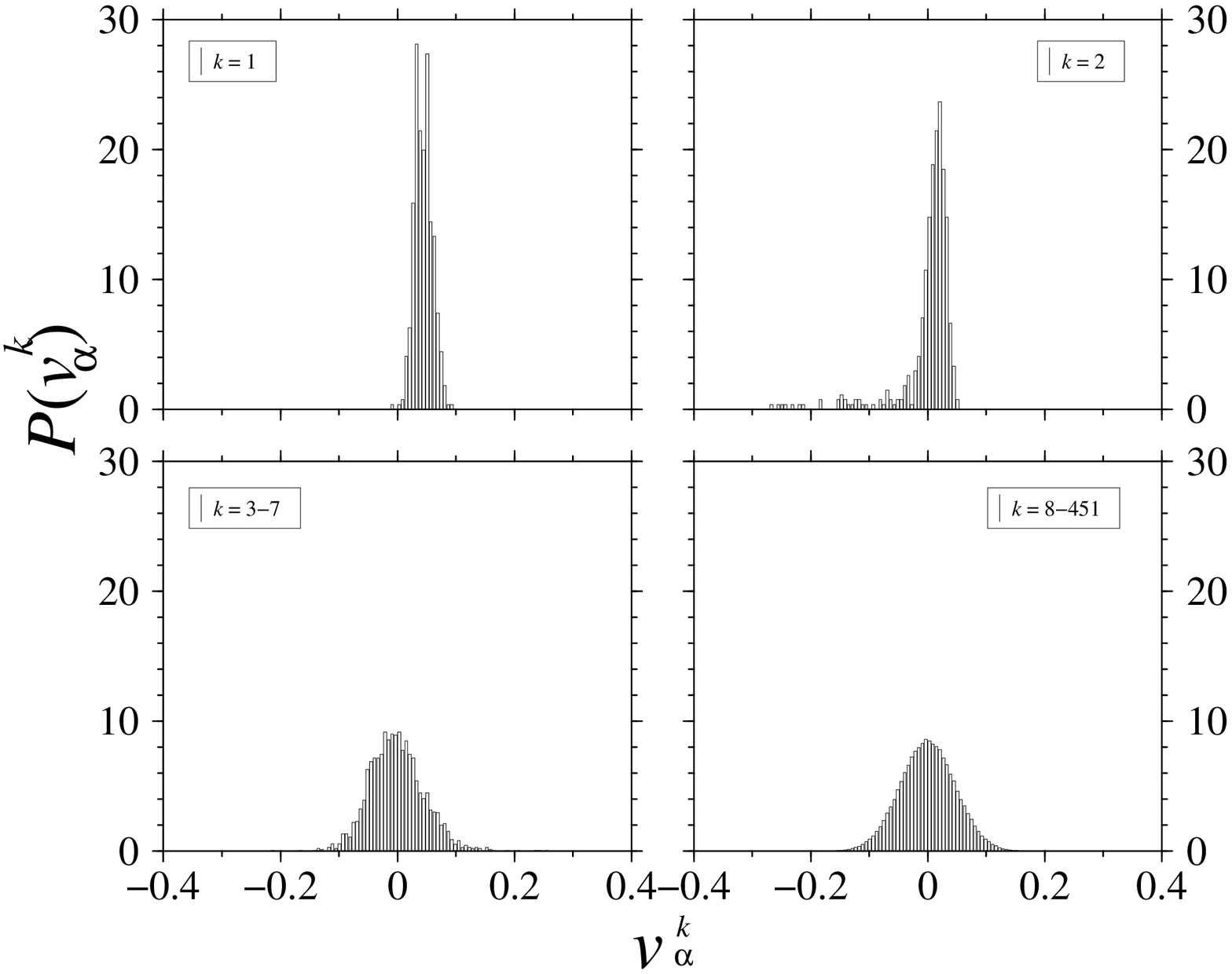}
\caption{Distributions of eigenvector components $v_{\alpha}^k$ for $k=1$,
$k=2$ and for the two sets $k=3-7$ and $k=8-451$ of eigenvectors. The
upper (lower) part corresponds to the returns (volatility) correlation
matrix.}
\label{fig:fig8}
\end{figure}

The first three $(k=1,2,3)$ analogously superposed volatility signals 
\begin{equation}
\nu_{{\lambda}_k}(t_i) = \sum_{\alpha=1}^N {\rm sign}(v^k_{\alpha}) 
\vert v^k_{\alpha} \vert^2 \vert g_{\alpha}(t_i) \vert.
\label{eq:nus}
\end{equation}
are shown in Fig~\ref{fig:fig7}. The first of them is associated with
the largest eigenvalue of the volatility correlation matrix and reflects 
the magnitude of average volatility as a function of time. The next two
$(k=2,3)$ constitute counterparts of the first two superposed return signals. 
Interestingly, this correspondence holds in reversed order, however.  

Another characteristics which carries some information about the stock
market dynamics is the probability distribution of the eigenvector components 
$v^k_{\alpha}$. Several relevant histograms, either for single eigenstates
or for a collection of them, are shown in Fig.~\ref{fig:fig8}.
In both cases, of the returns as well as of the volatility correlation 
matrices, the distributions of eigenvector components from the bulk 
of the spectrum agree very well with a Gaussian. 
In the transition region, illustrated here by the eigenstates 3-7,
some deviations can already be observed. Significantly different are the
distributions for the outliers. In the case of the returns correlation
matrix it is the second $(k=2)$ eigenvector whose distribution deviates
more from a Gaussian; its components are concentrated more at zero but,
at the same time, the tails of the distribution are thicker. This
indicates that fewer days $(\alpha$'s), but with a larger weight,
contribute to the signal seen at 14:30 than to
the one just after 17:00. The $k=1$ and $k=2$ eigenvector components of
the volatility correlation matrix are distributed asymmetrically relative
to zero, as consistent with the distribution of the entries of this matrix.
In this second case $(k=2)$, a long tail of the negative eigenvector 
components develops, which makes the corresponding superposed volatility
$\nu_{{\lambda}_2}(t_i)$ signal negative.       

\begin{figure}
\epsfxsize 8.0cm
\hspace{2.0cm}
\epsffile{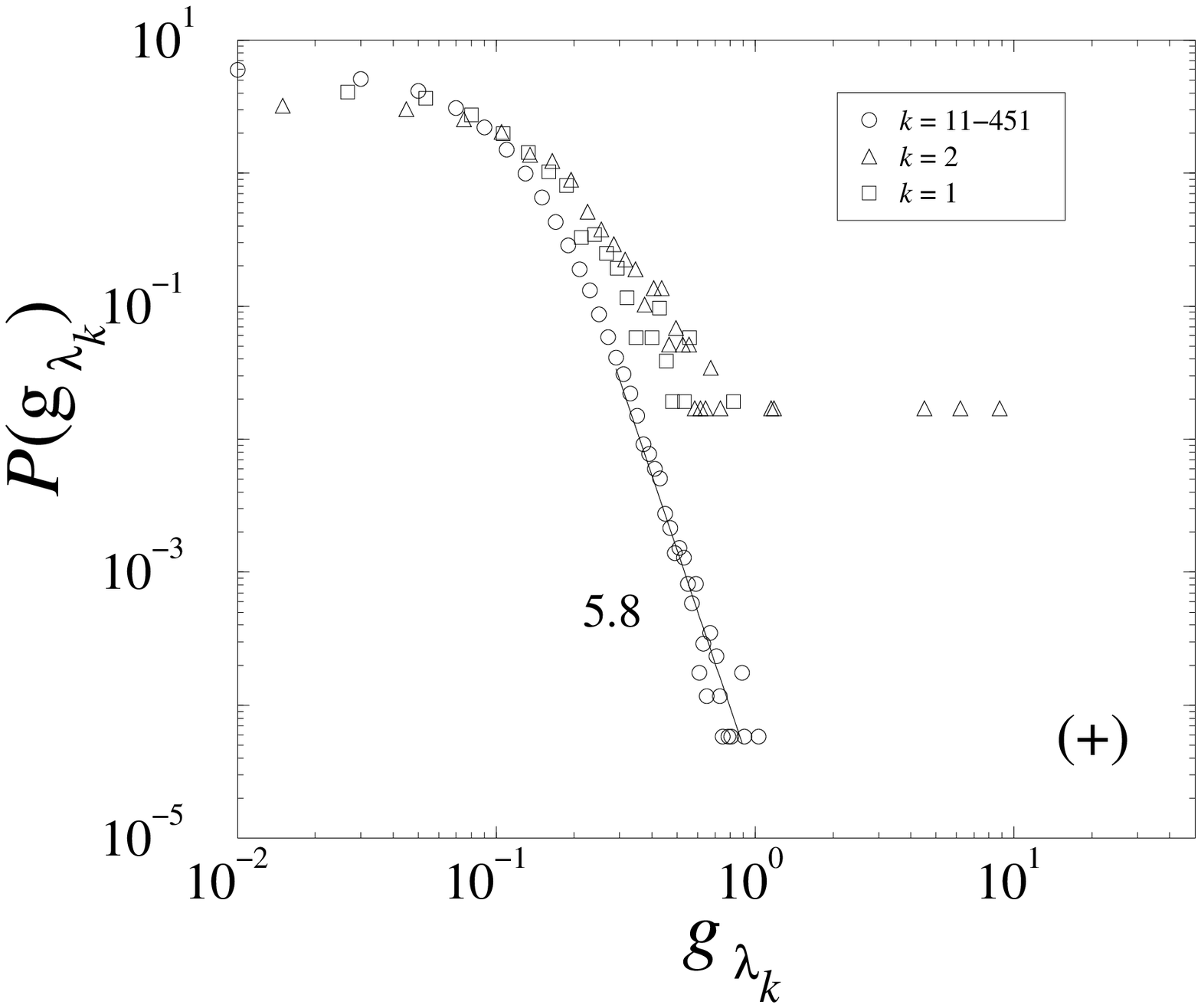}

\vspace{0.4cm}
\epsfxsize 8.0cm
\hspace{2.0cm}
\epsffile{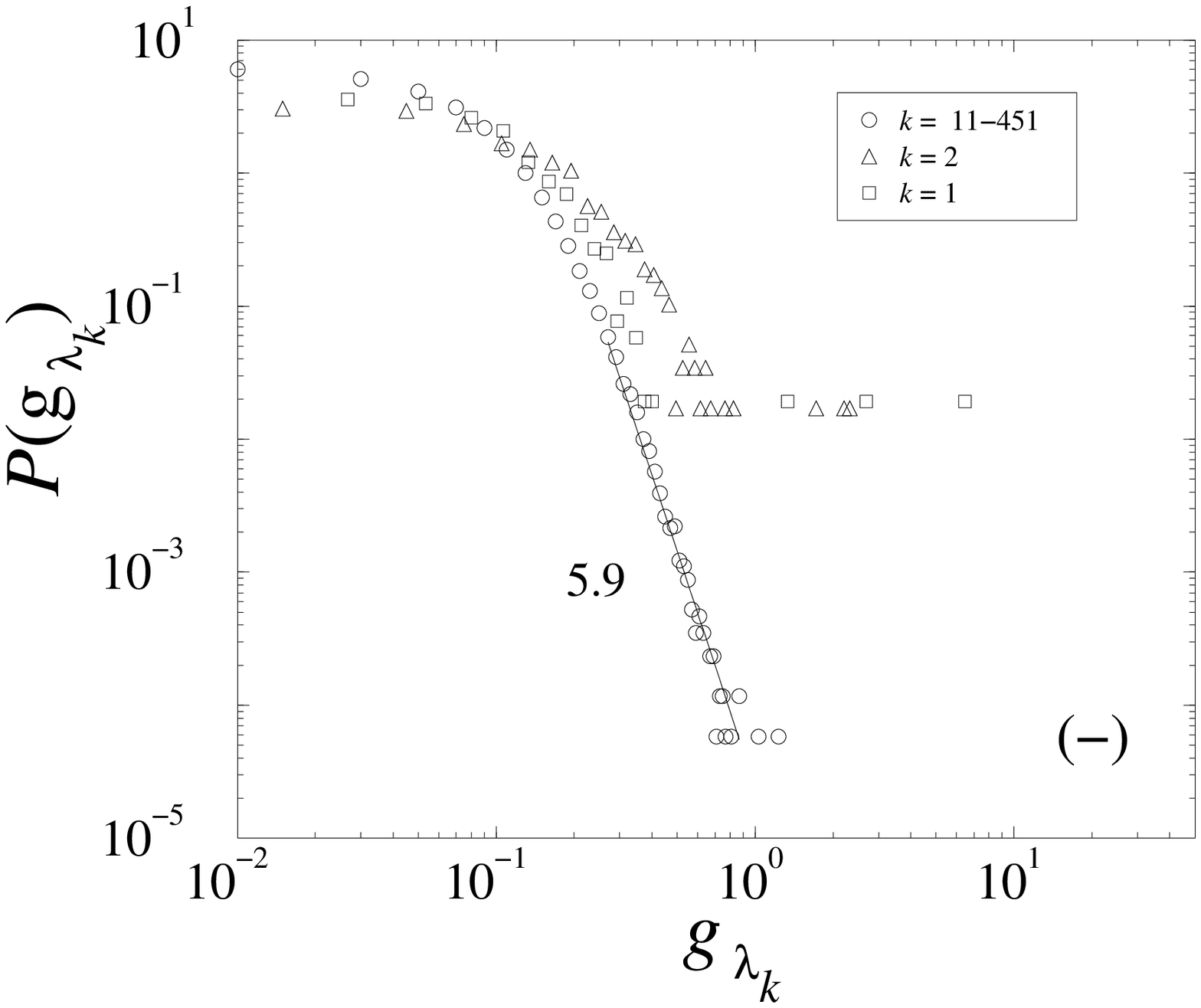}
\caption{Probability density functions of fluctuations of the 
superposed returns as expressed by the Eq.~(\ref{eq:gs}). Squares
correspond to $k=1$, triangles to $k=2$ and the circles to the average of
$k=11-451$. Both, positive (upper part) and negative (lower part) sides of
those distributions are shown.}
\label{fig:fig9}
\end{figure}

The above decomposition of the stock market dynamics allows also to shed
some more light on the issue of the probability distribution of price
changes~\cite{Stan}. That the nature of such changes is very complex
can be concluded by looking for instance at the probability distributions 
of fluctuations associated with different returns 'eigensignals' 
$g_{{\lambda}_k}(t_i)$. Some examples are shown in Fig.~\ref{fig:fig9}.
To quantify such characteristics like L\'evy stability or nonstability,
based on this analysis, would definitely be premature for many reasons. 
One is the statistics which is here too poor~\cite{Wero}.
What one however can clearly see is that the probability density
of fluctuations connected with the bulk of the spectrum drops down 
much faster than the ones connected with the more collective
$(k=1,2)$ eigenvectors. In the first case a power law index, of the order
of 5.8-5.9, can even be assigned, similar on both positive and negative
sides. The $k=1$ and $k=2$ signals trace however a completely different, 
much thicker tailed, distribution~\cite{nosign}. Their extreme events, 
which can be qualified as outliers~\cite{Lvov}, carry essentially the same 
features as the ones identified~\cite{Joha} in the Dow Jones draw downs
on much longer time scales.

\section {Summary}

The present study quantifies several characteristics relevant for
understanding the dynamics of the stock market time evolution.
One principal related issue is a question of how the market inefficiency 
manifests itself. For the Deutsche Aktienindex our study thus shows that 
on the time scales of up to one trading day there exist two 
well defined short periods of the spectacularly synchronous 
repeatable bursts of activity during the intraday trading between 9:03
and 17:10, a phenomenon somewhat in the sprit 
of an idea of marginally efficient markets~\cite{Zhan}.
On the other hand it turns out that generically the 
consecutive returns carry essentially no common information even when
probed with the frequency of 15 s.
The fluctuations associated with the so identified distinct components
are governed by the different laws which reflects an extreme 
complexity of the stock market dynamics.


\begin{thebibliography}{9}

\bibitem{Samu} P.A.~Samuelson, Industrial Management Rev. {\bf 6} (1965) 41

\bibitem{Lalo} L.~Laloux, P.~Cizeau, J-.P~Bouchaud, M.~Potters,
Phys.~Rev.~Lett. {\bf 83} (1999) 1467.

\bibitem{Pler} V.~Plerou, P.~Gopikrishnan, B.~Rosenow, L.A.N.~Amaral,
H.E.~Stanley, Phys.~Rev.~Lett. {\bf 83} (1999) 1471.

\bibitem{Dro1} S.~Dro\.zd\.z, F.~Gr\"ummer, A.Z.~G\'orski, F.~Ruf,
J.~Speth, Physica A {\bf 287} (2000) 440

\bibitem{Wign} E.P.~Wigner, Ann. Math. {\bf 53} (1951) 36.

\bibitem{Meht} M.L.~Mehta, Random Matrices, Academic Press, Boston, 1999

\bibitem{Dro2} S.~Dro\.zd\.z, F.~Gr\"ummer, F.~Ruf, J.~Speth,
Physica A {\bf 294} (2001) 226 

\bibitem{Camp} J.Y.~Campbell, A.W.~Lo, A.~Craig MacKinley, 
The Econometrics of Financial Markets, Princeton University Press, 
Princeton, NJ, 1997


\bibitem{Stan} R.N.~Mantegna, H.~Eugene Stanley, An Introduction to
Econophysics: Correlations and Complexity in Finance, University Press,
Cambridge, 2000

\bibitem{Kwap} J.~Kwapie\'n, S.~Dro\.zd\.z, A.A.~Ioannides,
Phys.~Rev.~E {\bf 62} (2000) 5557

\bibitem{Orme} P.~Ormerod, C.~Mounfield, Physica A {\bf 280} (2000) 497

\bibitem{Dro3} S.~Dro\.zd\.z, J.~Kwapie\'n, F.~Gr\"ummer, F.~Ruf,
J.~Speth, {\it Quantifying the dynamics of financial correlations},
cond-mat/0102402, Physica A {\bf 299} (2001) 144

\bibitem{data} H. Goeppl, Data from 'Karlsruher Kapitalmarktdatenbank (KKMDB)',
Institut f\"ur Entscheidungstheorie u. Unternehmensforschung,
Universit\"at Karlsruhe (TH)

\bibitem{Dro4} S.~Dro\.zd\.z, F.~Ruf, J.~Speth, M.~W\'ojcik,
Eur.~Phys.~J. {\bf B10} (1999) 589.

\bibitem{Edel} A.~Edelman, SIAM J.~Matrix Anal.~Appl. {\bf 9} (1988) 543;
\\ 
A.M.~Sengupta, P.P.~Mitra, Phys.~Rev.~E {\bf 60} (1999) 3389

\bibitem{Dro5} S.~Dro\.zd\.z, S.~Nishizaki, J.~Speth, M.~W\'ojcik,
Phys. Rev. E {\bf 57} (1998) 4016.

\bibitem{Dro6} S.~Dro\.zd\.z, F.~Gr\"ummer, F.~Ruf, J.~Speth, {\it
Dynamics of correlations in the stock market}, arXiv:cond-mat/0103606, \\
Empirical Science of Financial Fluctuations, Ed. H. Takayasu, Springer
Verlag, Tokyo, 2001, in press

\bibitem{Wero} R.~Weron, Int. J. Mod.Phys. C {\bf 12} (2001) 206

\bibitem{nosign} These observations remain largely unchanged  even if the
sign($v_{\alpha}^k$) in Eq.~(\ref{eq:gs}) is omitted. Then, only the
extreme events at 14:30, dominating $g_{\lambda_2}(t_i)$, are somewhat
reduced, but still remain outliers.

\bibitem{Lvov} V.S.~Lvov, A.~Pomyalov, I.~Procaccia, 
Phys. Rev. E {\bf 63} (2001) 056118

\bibitem{Joha} A.~Johansen, D.~Sornette, Eur. Phys. J. B {\bf 9} (1999) 167

\bibitem{Zhan} Y.-C.~Zhang, Physica A {\bf 269} (1999) 30

\end{thebibliography}
\end{document}